\documentclass[letterpaper,twocolumn,10pt]{article}

\usepackage{zhanggroup}
\usepackage[absolute]{textpos}
\usepackage{amsmath}
\usepackage{tikz,eso-pic}
\usepackage{amsfonts}
\usepackage{xspace} 
\usepackage{mathrsfs}
\usepackage{amsmath}
\usepackage{amsthm}
\usepackage{subcaption}
\usepackage{booktabs}
\usepackage{multirow}
\usepackage{graphicx}
\usepackage{flushend}
\usepackage{url}
\usepackage{xurl}
\usepackage{caption, subcaption}
\usepackage{float}
\usepackage[normalem]{ulem}
\usepackage{hyperref}
\hypersetup{
  colorlinks,
  linkcolor={blue!70!green},
  citecolor={green!70!blue},
  urlcolor={orange!70!red}
}
\usepackage[breakable]{tcolorbox}
\newtcolorbox{cvbox}[1][]{
    after skip=8mm,
    title=#1,
    breakable = true,
    fonttitle=\sffamily\bfseries,
    coltitle=white,
    colbacktitle=gray!100,   
    titlerule= 0pt,         
    overlay={%
        \ifcase\tcbsegmentstate
        \or%
        \else%
        \fi%
    }
    colback = gray,         
    colframe = black!75     
    }
    
\newcommand{\mypara}[1]{\smallskip\noindent{\bf {#1}.}\xspace}

\begin{document}

\date{}

\title{Excessive Reasoning Attack on Reasoning LLMs}

\author{
{\rm Wai Man Si},\ \ \
{\rm Mingjie Li},\ \ \
{\rm Michael Backes},\ \ \
{\rm Yang Zhang}
\\
\\
\textit{CISPA Helmholtz Center for Information Security}\ \ \
}

\maketitle

\begin{abstract}

Recent reasoning large language models (LLMs), such as OpenAI o1 and DeepSeek-R1, exhibit strong performance on complex tasks through test-time inference scaling.
However, prior studies have shown that these models often incur significant computational costs due to excessive reasoning, such as frequent switching between reasoning trajectories (e.g., underthinking) or redundant reasoning on simple questions (e.g., overthinking).
In this work, we expose a novel threat: adversarial inputs can be crafted to exploit excessive reasoning behaviors and substantially increase computational overhead without compromising model utility.
Therefore, we propose a novel loss framework consisting of three components:
(1) Priority Cross-Entropy Loss, a modification of the standard cross-entropy objective that emphasizes key tokens by leveraging the autoregressive nature of LMs;
(2) Excessive Reasoning Loss, which encourages the model to initiate additional reasoning paths during inference; and
(3) Delayed Termination Loss, which is designed to extend the reasoning process and defer the generation of final outputs.
We optimize and evaluate our attack for the GSM8K and ORCA datasets on DeepSeek-R1-Distill-LLaMA and DeepSeek-R1-Distill-Qwen. 
Empirical results demonstrate a 3x to 9x increase in reasoning length with comparable utility performance.
Furthermore, our crafted adversarial inputs exhibit transferability, inducing computational overhead in o3-mini, o1-mini, DeepSeek-R1, and QWQ models.

\end{abstract}

\section{Introduction}
\label{sec:intro}

Inference-time scaling has emerged as a critical technique for enhancing the reasoning capabilities of LLMs~\cite{WWSBIXCLZ22, WWSLCNCZ23, BBKGPGGLNNH24, YYZCXZGC24}.
Methods such as Chain-of-Thought (CoT)~\cite{WWSBIXCLZ22} explicitly guide models through intermediate reasoning steps, while recent models like DeepSeek-R1~\cite{DGYZSZXZMWBZYWWGSLGLXWWFLZDZRDCJLLDLHCLZBXWDXGQLGLWCYQLCNLCDHGGHYWZZWZXXZZTLWLTHZWCDGZPWCJCLZCYWYZPL25} and QWQ~\cite{qwq32b} incorporate reasoning capabilities implicitly during training.
However, recent studies have shown that these models often suffer from excessive reasoning behaviors, such as frequent shifts in reasoning strategies or redundant processing, which can lead to substantial computational overhead~\cite{WLXLCHSYLZWTMY25, CXLHPYSLZZWTMY24}.
These inefficiencies introduce a novel security risk: the attacker can craft adversarial inputs to exploit them, significantly inflating inference-time resource usage.

Prior work has explored related threats in both language models and vision-language models (VLMs).
For example, Sponge examples~\cite{SZBPMA21} increase computational costs by maximizing activation norms, while the NICGSlowdown attack~\cite{CSHLY22} manipulates token logits to delay output generation. 
Similarly, Gao et al.~\cite{GBGXTLL24} introduce verbose images to impose high inference latency and computational burden specifically on VLMs. 
More recently, Kumar et al.~\cite{KRNKIHB25} propose the Overthink attack, which adopts an indirect prompt injection strategy and inserts a decoy to external resources.
This compels the model to allocate additional reasoning resources toward solving an intermediary task before addressing the primary query.
In contrast, our method directly perturbs the input to elicit excessive reasoning behavior, increasing computational overhead without degrading task performance or requiring external content.
Moreover, our attack aligns with the Model Denial of Service (MDoS) threat as defined by OWASP, wherein adversarial inputs lead to resource exhaustion, degrading system responsiveness and service availability for other users.\footnote{\url{https://genai.owasp.org/llmrisk2023-24/llm04-model-denial-of-service/}}

In this work, we introduce the first adversarial attack designed to exploit the reasoning inefficiencies in reasoning LLMs, thus inducing excessive computation during inference. 
Our approach constructs adversarial suffixes that prompt the model to engage in extended reasoning without compromising model utility.
To optimize these suffixes, we propose three novel loss functions that encourage such reasoning behavior:
\begin{itemize}
    \item \textbf{Priority Cross-Entropy Loss} prioritizes key tokens while masking less informative ones to enhance optimization efficiency. 
    This loss leverages the autoregressive nature of LM to enable more targeted and effective gradient updates.
    \item \textbf{Excessive Reasoning Loss} increases the likelihood of branched or recursive reasoning, leading to greater computational overhead.
    \item \textbf{Delayed Termination Loss} encourages the model to defer the termination of reasoning and answer generation.
\end{itemize}

We optimize and evaluate our attacks for the GSM8K~\cite{CKBCJKPTHNHS21} and ORCA~\cite{MKRA24} datasets on DeepSeek-R1-Distill-Llama and DeepSeek-R1-Distill-Qwen.
Our attacks consistently increase the reasoning length by over 3x to 9x using only 10 crafted adversarial tokens.
Moreover, our attack demonstrates strong transferability across models on commercial platforms, including OpenAI o1-mini and o3-mini~\cite{JKLRELHMBCIKPNPWTBKSVDKMAJNZGRSBMMHBHMELBHLBVSZKOHFCRKLSDFMRTLOFZWPCMWRMWSRLTSLSSPCZBLSBSiBRLCKOOGA24}, DeepSeek-R1, and QWQ, suggesting a broader vulnerability among reasoning-optimized LLMs.
These findings expose an underexplored issue.
While such models are proficient in reasoning, they remain susceptible to targeted manipulations that exploit their reasoning mechanisms to induce significant computational overhead.
Our results underscore the urgent need for inference-time defenses that can detect and mitigate excessive reasoning triggered by adversarial prompts, particularly in real-world deployments.

\section{Methodology}
\label{sec:method}

This section introduces our adversarial attack framework, which aims to increase the computational overhead of reasoning LLMs by inducing excessive reasoning behavior. 
We first formalize the threat model, then describe the procedure for generating target outputs.
Finally, we detail the loss functions that guide the optimization.

\subsection{Threat Model}

The primary objective of our attack is to generate inputs that compel the model to extend the reasoning processes as long as possible, thus significantly increasing the computational cost at inference time.
Also, this manipulation should preserve the model's utility to avoid suspicion. 
Similar to prior work~\cite{CNCJGKITS23, ZWKF23, GBGXTLL24, BSAP22}, we assume a white-box scenario in which the attacker has complete access to the model’s architecture, parameters, and gradients.

Following~\cite {CNCJGKITS23}, we consider two primary use cases for our attack.
In the first use case, a malicious user intentionally induces excessive computational load, degrading overall system performance and diminishing service quality for other users, akin to a DoS attack.
In the second use case, a benign user queries the model within an autonomous system that processes untrusted third-party data (e.g., crafted adversarial data), resulting in significantly higher costs (e.g., money) than expected.
As we later demonstrate, these crafted adversarial inputs exhibit strong transferability across different models.

\subsection{Target Output Generation}

Constructing an effective adversarial suffix requires defining a target output that can guide the optimization process. 
Prior work has adopted similar strategies in various contexts.
For instance, ATA~\cite{GZCMGKKS24} uses the fixed string ``Sorry, I’m unable to answer the question'' to mislead the model into generating incorrect answers, while Zou et al.~\cite{ZWKF23} target phrases such as ``Sure, I can...'' to bypass safety mechanisms and elicit unsafe behavior.

For our attack, a straightforward strategy is to sample multiple outputs from the target model and select the longest one as the optimization target.
However, we find that this approach often fails to produce outputs of sufficient length. 
Another option is to use reasoning-inducing prompts such as CoT, which are designed to elicit a step-by-step reasoning path. 
Although promising, our experiments show that CoT prompts do not consistently generate longer outputs across various models and datasets.

To further increase target output length, we adopt DSPy~\cite{KSMZSVHSJMMZP24}, a recent prompt optimization framework that iteratively refines instructions to better satisfy a given objective.
Specifically, we use a DSPy optimizer to refine CoT prompts on a small dataset with the goal of maximizing output length.
The resulting optimized CoT prompts elicit substantially longer responses from the target model and serve as effective targets for crafting adversarial examples.
We include the optimized prompt in App.~\ref{box:dspy_prompt} and report output length statistics for different prompting strategies in Table~\ref{tab:target_stat}.

\begin{figure*}[t]
  \centering
  \includegraphics[width=.9\textwidth]{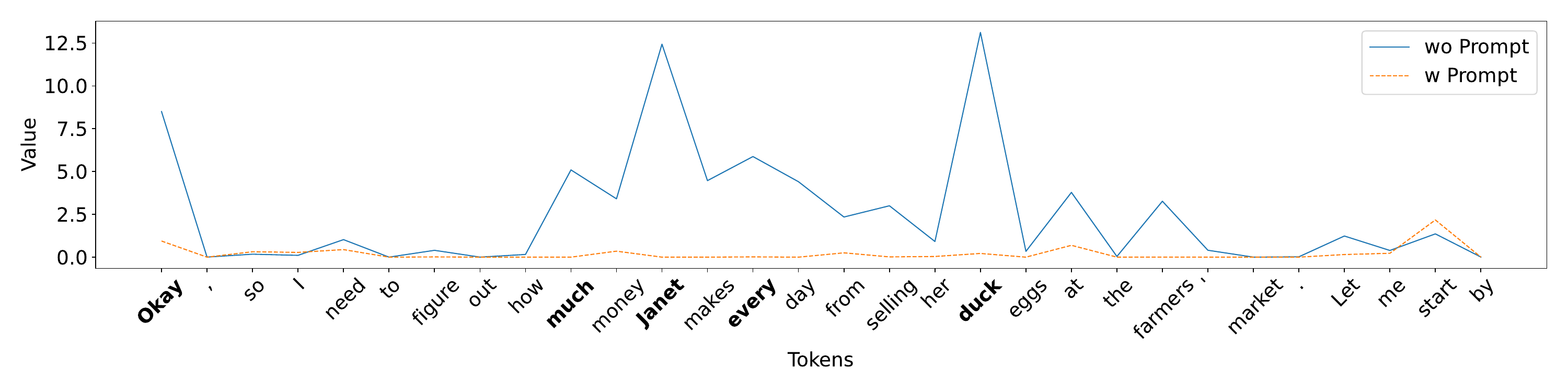}
  \caption{The perplexity of a reasoning sample in the output with and without the prompt.
  Bold tokens are assigned 1 in the mask.}
  \label{fig:ppl_study}
\end{figure*}

\subsection{Loss Design}

To craft adversarial suffixes that trigger excessive reasoning behavior, we propose a composite loss function consisting of three components: Priority Cross-Entropy (PCE) Loss, Excessive Reasoning (ER) Loss, and Delayed Termination (DT) Loss. 
PCE Loss is designed to reduce optimization difficulty, while ER Loss and DT Loss target distinct aspects of the reasoning process.
We detail each component below.

\mypara{Priority Cross-Entropy Loss}
Traditional adversarial attacks on LMs typically optimize a cross-entropy loss to maximize the likelihood of generating a specific target sequence (e.g., ``Sure, I can ...'').
Formally, given an input token sequence \( x = \{w_1, w_2, \dots, w_n\} \), the probability of generating the next token \( w_{n+1} \) is defined as:
\begin{equation}
    p(w_{n+1}) = p(w_{n+1} \mid w_1, w_2, \dots, w_n).
\end{equation}

Accordingly, the standard cross-entropy loss used to maximize the likelihood of a target sequence \( y \), conditioned on a base input $x$ and an adversarial suffix $x'$, is defined as:
\begin{equation}
    \mathcal{L}_{\text{CE}} = -\frac{1}{|y|} \sum_{t=1}^{|y|} \log p(y_t \mid \{x, x'\}, y_{<t}).
\end{equation}

In typical adversarial settings, the target sequence \( y \) is relatively short, often fewer than 10 tokens.
However, to trigger excessive reasoning behavior, we must construct much longer targets (e.g., over 1,000 tokens).
Uniformly optimizing over such long sequences is computationally inefficient, as many tokens (e.g., ``the'' and ``I'') can be accurately generated even without the prompt, due to statistical priors learned during pretraining.
To investigate this effect, we analyze the loss distribution of a target sequence with and without the input prompt.
As shown in Fig.~\ref{fig:ppl_study}, our analysis reveals that only a small subset of tokens exhibits a significant increase in loss when the prompt is removed.
This observation supports our hypothesis that informativeness is not uniformly distributed across tokens and that only a subset is highly dependent on the input prompt.

Building on this insight, we introduce a token-level importance mask that emphasizes tokens the model considers informative, thereby improving optimization efficiency.
Specifically, for each target token $y_t$, we compute an importance score as the difference in log-probabilities with and without the input prompt:
\begin{equation}
    \text{S}_t = \log p(y_t \mid y_{<t}) - \log p(y_t \mid x, y_{<t}).
\end{equation}
This score captures the degree to which each token’s prediction depends on the presence of the prompt.
We then construct a binary mask $\mathcal{M}$ by selecting the top $K$ \% of tokens with the highest importance scores and assigning them a value of 1, while masking out the remaining tokens (assigned 0). 
The resulting PCE Loss is defined as:
\begin{equation}
    \mathcal{L}_{\text{PCE}} = - \frac{1}{|y|} \sum_{t=1}^{|y|} \mathcal{M}_{t} \cdot \log p(y_t \mid \{x, x'\}, y_{<t}).
\end{equation}
By selectively focusing on prompt-sensitive tokens, this loss function enhances optimization efficiency and more effectively encourages the model to generate extended reasoning sequences during inference.

\mypara{Excessive Reasoning Loss}  
Prior work~\cite{WLXLCHSYLZWTMY25, CXLHPYSLZZWTMY24} has shown that LLMs trained for explicit reasoning often produce extended reasoning.
In such cases, certain tokens, such as \textit{``Wait''} and \textit{``Alternative''}, frequently occur in these sequences, signaling branching or recursive reasoning steps.
To exploit this behavior, we aim to increase the likelihood of generating such tokens during the reasoning.
While constructing a manual list of indicative tokens is feasible, it is limited in scalability and generalization.
Instead, we adopt a data-driven approach to automatically identify reasoning-associated tokens.
As demonstrated in our ablation study (Sec.~\ref{sec:ablation}), this approach uncovers influential tokens that would likely be overlooked through manual inspection, underscoring the efficacy of our method.

Concretely, we extract the top \(n\) most frequent tokens that appear in the first two positions of sentences generated during the Target Output Generation phase.
These tokens are hypothesized to play a critical role in initiating new reasoning trajectories. 
Let \( \mathcal{T} \) denote the resulting set of high-impact tokens.
To promote their occurrence during generation, we define the Excessive Reasoning (ER) Loss as:
\begin{equation}
    \mathcal{L}_{\text{ER}} = - \frac{1}{|y|} \sum_{t=1}^{|y|} \sum_{v \in \mathcal{T}} \log p(y_t = v \mid \{x, x'\}, y_{<t}).
\end{equation}

This objective increases the likelihood of generating tokens associated with recursive or exploratory reasoning, thereby inducing longer and more computationally intensive reasoning sequences.

\mypara{Delayed Termination Loss}  
In many reasoning LLMs, the generation process typically begins with intermediate reasoning steps, which conclude with a designated end-of-thinking (EOT) token (e.g., \texttt{</think>}).
Then, the model would generate an answer conclusion terminated by an end-of-sequence (EOS) token (e.g., \texttt{<eos>}).
To prolong both the reasoning and answer conclusion phases, we aim to reduce the model’s tendency to emit these termination tokens during decoding.
However, due to the stochastic nature of autoregressive generation, the precise timestep at which these tokens appear is not fixed.
To address this, we adopt a strategy from prior work~\cite{CSHLY22, GBGXTLL24}, which minimizes the likelihood of generating termination tokens across all positions in the output sequence:
\begin{align}
    \mathcal{L}_{\text{DT}} &= \frac{1}{|y|} \sum_{t=1}^{|y|} \Big[p(y_t = \text{EOS} \mid \{x, x'\}, y_{<t}) \nonumber \\
    &\quad +\; p(y_t = \text{EOT} \mid \{x, x'\}, y_{<t}) \Big].
\end{align}

This objective discourages premature termination, encouraging the model to continue generating extended reasoning and answer conclusions before finalizing its output.

\subsection{Optimization}

Optimizing adversarial suffixes in the text domain presents a unique challenge due to the discrete nature of language. 
Unlike continuous domains (e.g., images), where gradients can be directly applied to pixel values, LMs operate on sequences of discrete tokens drawn from a fixed vocabulary. 
As a result, standard gradient-based optimization techniques cannot be directly applied to manipulate individual tokens.

To address this, we adopt the Greedy Coordinate Gradient-based Search (GCG) framework~\cite{ZWKF23}, which has demonstrated strong performance in adversarial text generation. 
GCG linearizes the loss landscape by computing gradients with respect to input embeddings and identifying substitutions that are most likely to improve the loss.
Specifically, for a given token position \( i \) in the suffix, we compute the gradient of the loss with respect to its embedding and search for the token $x_i'$ that maximally improves the objective. Formally:
\begin{equation}
    x_i' = \arg\max_{w \in V} \left\langle \nabla_{e(x_i)} \mathcal{L},\ e(w) - e(x_i) \right\rangle,
\end{equation}
where \( \nabla_{e(x_i)} \mathcal{L} \) denotes the gradient of the loss with respect to the embedding of token \( x_i \), and \( e(w) \) is the embedding of candidate token \( w \). 
This inner product quantifies the expected gain from substituting \( x_i \) with \( w \), and the best candidate is selected greedily.
Our overall training objective combines the three loss components introduced previously:
\begin{equation}
    \mathcal{L} = \alpha \cdot \mathcal{L}_{\text{PCE}} + \beta \cdot \mathcal{L}_{\text{ER}} + \gamma \cdot \mathcal{L}_{\text{DT}}.
\end{equation}

In this work, we adopt a fixed-length suffix-based strategy in which a predetermined number of tokens are appended to the end of the original prompt. 
Each token in the suffix is iteratively updated using GCG to minimize the combined loss.
Although this paper focuses on crafting adversarial suffixes, it is important to note that our approach is \textit{method-agnostic} and can be adapted to various adversarial paradigms.
For instance, alternative strategies such as character-level perturbations (e.g., typos) can also be incorporated, as shown in prior work~\cite{GZCMGKKS24}.
This flexible framework facilitates the efficient generation of adversarial inputs tailored to different attack objectives and constraints.

\section{Experiments}

\subsection{Experimental Setups}
\label{sec:exp_setups}

\mypara{Models and Datasets}  
We optimize adversarial suffixes and evaluate them on two reasoning LLMs: DeepSeek-R1-distill-LLaMA-8B and DeepSeek-R1-distill-Qwen-7B.\footnote{For simplicity, we omit the prefix DeepSeek-R1-distill throughout the remainder of the paper.}
Both models are distilled variants of DeepSeek-R1 and demonstrate strong performance on complex reasoning tasks.
We report results under two decoding strategies: greedy decoding and sampling decoding. 
For sampling, we set the temperature to 0.6 and apply nucleus sampling with top-$p = 0.95$.
To assess cross-model transferability, we additionally evaluate the attack on larger-scale models, including o1-mini, o3-mini, DeepSeek-R1, and QWQ-32B, using their respective default decoding settings. 
Specifically, we interact with o1-mini and o3-mini via the OpenAI API, and with DeepSeek-R1 and QWQ-32B via the Baidu Cloud API, to simulate real-world deployment conditions.
Our evaluation is conducted on two widely used mathematical reasoning benchmarks: GSM8K~\cite{CKBCJKPTHNHS21} and ORCA~\cite{MKRA24}. 
For each dataset, we randomly sample 50 examples for both optimization and evaluation.

\mypara{Attack Setup} 
For target output generation, we employ the COPRO optimizer to construct prompts that induce extended reasoning trajectories. 
Specifically, we use 10 training examples from the GSM8K dataset for prompt optimization and evaluate the resulting prompt on a separate set of 10 randomly selected test samples.  
Due to computational constraints, we restrict target outputs to a maximum length of 3,000 tokens.
For the PCE Loss, we set the token selection threshold \( K = 1 \), and for the ER Loss, we use \( n = 5 \).  
The overall loss function combines the three components using the following weighting coefficients: \( \alpha = 1 \), \( \beta = 50 \), and \( \gamma = 1 \).  
We fix the length of the adversarial suffix to 10 tokens.  
During optimization, we apply the GCG algorithm for 1,000 steps per input. 
The candidate pool size is set to 64, and at each step, the top 64 candidate tokens are retained.

\mypara{Evaluation Metrics}  
To evaluate the effectiveness of our adversarial attack, we consider three primary metrics: (1) output sequence length (in tokens), (2) inference latency (in seconds), and (3) energy consumption (in Joules).
Energy usage is measured using the NVIDIA Management Library (NVML), following the methodology introduced by Shumailov et al.~\cite{SZBPMA21}.
To ensure consistency and fair comparison, all inference is performed using the HuggingFace pipeline~\cite{huggingface} on a single hardware (NVIDIA A100 80GB).
Each inference is repeated three times to reduce the impact of runtime variability.
To assess model utility, we extract final answers from the generated outputs using ``Meta-Llama-3.1-8B-Instruct'', and compute accuracy by comparing the extracted answers against ground-truth labels. 
The exact prompt used for extraction is provided in App.~\ref{box:acc_prompt}.

\mypara{Baselines}  
To evaluate the effectiveness of our attack, we compare it against several baseline prompting strategies:
\begin{itemize}
    \item \textbf{Random}: A suffix composed of 10 randomly sampled tokens.
    \item \textbf{Standard CoT~\cite{WWSBIXCLZ22}}: A widely used CoT prompt that appends the phrase ``Let's think step by step.''
    \item \textbf{CatAttack~\cite{RRTYBMZR25}}: A prompt-based adversarial strategy that appends the distractor statement ``Interesting fact: cats sleep most of their lives,'' which has been shown to induce incorrect reasoning outputs.
\end{itemize}

\begin{table*}[t]
\centering
\resizebox{.9\linewidth}{!}{
\begin{tabular}{llcccccc|cccccc}
\toprule
 & \multicolumn{1}{c}{} & \multicolumn{6}{c|}{GSM8K} & \multicolumn{6}{c}{ORCA} \\
Models & \multicolumn{1}{c}{Methods} & \multicolumn{1}{c}{Rea} & \multicolumn{1}{c}{Ans} & \multicolumn{1}{c}{Full} & \multicolumn{1}{c}{Lat} & \multicolumn{1}{c}{Ent} & \multicolumn{1}{c|}{Acc} & \multicolumn{1}{c}{Rea} & \multicolumn{1}{c}{Ans} & \multicolumn{1}{c}{Full} & \multicolumn{1}{c}{Lat} & \multicolumn{1}{c}{Ent} & \multicolumn{1}{c}{Acc} \\
\midrule
\multirow{5}{*}{LLaMA} & Original & 574 & \textbf{266} & 839 & 22.2 & 4712 & 70\% & 344 & \textbf{259} & 603 & 14.7 & 2789 & \textbf{82\%} \\
  & Random & 496 & 232 & 729 & 19.7 & 3959 & 76\% & 338 & 233 & 571 & 15.3 & 3013 & 78\% \\
 & CoT & 447 & 264 & 711 & 19.0 & 3571 & 74\% & 440 & 228 & 668 & 17.6 & 3570 & 76\% \\
 & CatAttack & 668 & 239 & 907 & 24.3 & 4628 & 72\% & 499 & 213 & 712 & 19.3 & 4366 & \textbf{82\%} \\
 & Ours & \textbf{1914} & 160 & \textbf{2074} & \textbf{54.9} & \textbf{12827} & \textbf{92\%} & \textbf{1575} & 167 & \textbf{1743} & \textbf{47.2} & \textbf{9929} & 80\% \\
 \midrule
\multirow{5}{*}{Qwen} & Original & 169 & \textbf{310} & 479 & 11.9 & 2535 & 82\% & 379 & \textbf{248} & 626 & 15.6 & 2910 & 86\% \\
 & Random & 167 & 295 & 461 & 11.3 & 2171 & 78\% & 532 & 234 & 766 & 18.5 & 4040 & 82\% \\
 & CoT & 159 & 294 & 453 & 11.2 & 2097 & 82\% & 527 & 225 & 752 & 18.7 & 3505 & 86\% \\
 & CatAttack & 237 & 282 & 519 & 12.8 & 2498 & 84\% & 531 & 220 & 750 & 18.4 & 4034 & 84\% \\
 & Ours & \textbf{1531} & 193 & \textbf{1724} & \textbf{42.4} & \textbf{8188} & \textbf{88\%} & \textbf{1459} & 166 & \textbf{1624} & \textbf{39.6} & \textbf{9155} & \textbf{88\%} \\
\bottomrule
\end{tabular}
}
\caption{
The token length for reasoning (Rea), answer (Ans), and full output (Full); inference latency (Lat, in seconds); energy consumption (Ene, in joules); and task accuracy (Acc). 
Experimental results across methods under greedy decoding.
\textbf{Bold} indicates the best result.
}
\label{tab:main_greedy}
\end{table*}

\begin{table*}[t]
\centering
\resizebox{.9\linewidth}{!}{
\begin{tabular}{llcccccc|cccccc}
\toprule
 & \multicolumn{1}{c}{} & \multicolumn{6}{c|}{GSM8K} & \multicolumn{6}{c}{ORCA} \\
Models & \multicolumn{1}{c}{Methods} & \multicolumn{1}{c}{Rea} & \multicolumn{1}{c}{Ans} & \multicolumn{1}{c}{Full} & \multicolumn{1}{c}{Lat} & \multicolumn{1}{c}{Ent} & \multicolumn{1}{c|}{Acc} & \multicolumn{1}{c}{Rea} & \multicolumn{1}{c}{Ans} & \multicolumn{1}{c}{Full} & \multicolumn{1}{c}{Lat} & \multicolumn{1}{c}{Ent} & \multicolumn{1}{c}{Acc} \\
\midrule
\multirow{5}{*}{LLaMA} & Original & 476 & \textbf{280} & 757 & 68.5 & 7713 & 75\% & 402 & \textbf{266} & 668 & 75.9 & 9148 & 80\% \\
  & Random & 528 & 257 & 785 & 64.8 & 9818 & 77\% & 475 & 224 & 700 & 81.4 & 10680 & 82\% \\
 & CoT & 401 & 270 & 671 & 54.0 & 7928 & 72\% & 493 & 244 & 737 & 80.6 & 9689 & 81\% \\
 & CatAttack & 556 & 257 & 812 & 71.5 & 9778 & 76\% & 550 & 248 & 798 & 86.7 & 10490 & 81\% \\
 & Ours & \textbf{1437} & 204 & \textbf{1641} & \textbf{197.0} & \textbf{21228} & \textbf{90\%} & \textbf{1425} & 206 & \textbf{1631} & \textbf{178.2} & \textbf{19243} & \textbf{87\%} \\
 \midrule
\multirow{5}{*}{Qwen} & Original & 176 & \textbf{308} & 484 & 16.0 & 3799 & 85\% & 293 & \textbf{274} & 567 & 33.1 & 5722 & \textbf{87\%} \\
 & Random & 187 & 295 & 482 & 17.4 & 3398 & 81\% & 432 & 253 & 686 & 50.0 & 6458 & 83\% \\
 & CoT & 221 & 296 & 518 & 23.4 & 4245 & 84\% & 501 & 254 & 755 & 54.5 & 7353 & 86\% \\
 & CatAttack & 345 & 277 & 622 & 37.0 & 5996 & 82\% & 452 & 250 & 701 & 51.2 & 7538 & 84\% \\
 & Ours & \textbf{1479} & 217 & \textbf{1696} & \textbf{149.1} & \textbf{16031} & \textbf{91\%} & \textbf{1238} & 183 & \textbf{1421} & \textbf{111.2} & \textbf{14747} & \textbf{87\%} \\
 \bottomrule
\end{tabular}
}
\caption{Experimental results across methods under sampling decoding.}
\label{tab:main_sample}
\end{table*}

\subsection{Main Results}  
\label{sec:result}

\mypara{Performance}
We evaluate the effectiveness of our attack using six metrics: reasoning token length (Rea), answer token length (Ans), total output length (Full), inference latency (Lat), energy consumption (Ene), and task accuracy (Acc). 
Evaluations are conducted under both greedy and sampling-based decoding strategies.
As shown in Table~\ref{tab:main_greedy} and Table~\ref{tab:main_sample}, our adversarial suffix substantially increases computational overhead while preserving task accuracy across all settings.
For example, our attack causes LLaMA to generate significantly longer outputs on the GSM8K dataset with greedy decoding, increasing the average reasoning length by 3x from 574 to 1,914 tokens.
This is accompanied by a corresponding increase in energy consumption (from 4,712J to 12,827J) and latency (from 22.2s to 54.9s).
A similar trend is observed for Qwen, where the average reasoning length increases by 9x, demonstrating the effectiveness of our attack across different model architectures.
Under sampling-based decoding, the attack remains robust.
The reasoning length increases by 3x for LLaMA and 8x for Qwen on GSM8K, with similar results observed on the ORCA dataset.

In comparison, baseline prompting methods generally induce relatively short reasoning.
For example, CoT prompts produce shorter outputs than our adversarial prompt on LLaMA for GSM8K under greedy decoding (711 vs. 2,074 tokens), indicating the limitations of standard methods in eliciting excessive reasoning behavior. 
More broadly, our results suggest that reasoning LLMs are resistant to short prompting, as neither CoT nor CatAttack reliably trigger long reasoning.
Interestingly, we observe a consistent inverse correlation between the lengths of reasoning and answer segments.
We hypothesize that as the model allocates more capacity to the reasoning phase, the corresponding answer portion becomes shorter.
Importantly, the increase in reasoning length does not degrade task accuracy; in many cases, it correlates with improved performance.
This suggests that excessive reasoning may enhance the model’s problem-solving capabilities. 
Thus, our attack exhibits a dual effect: it exposes a vulnerability in inference-time efficiency while potentially enhancing the model’s reasoning capabilities.
As a result, such adversarial behaviors may evade detection by standard evaluation metrics that focus solely on output correctness, highlighting the need for more comprehensive evaluation frameworks.

\begin{figure}[t]
\centering
\begin{subfigure}{\columnwidth}
    \includegraphics[width=.9\linewidth]{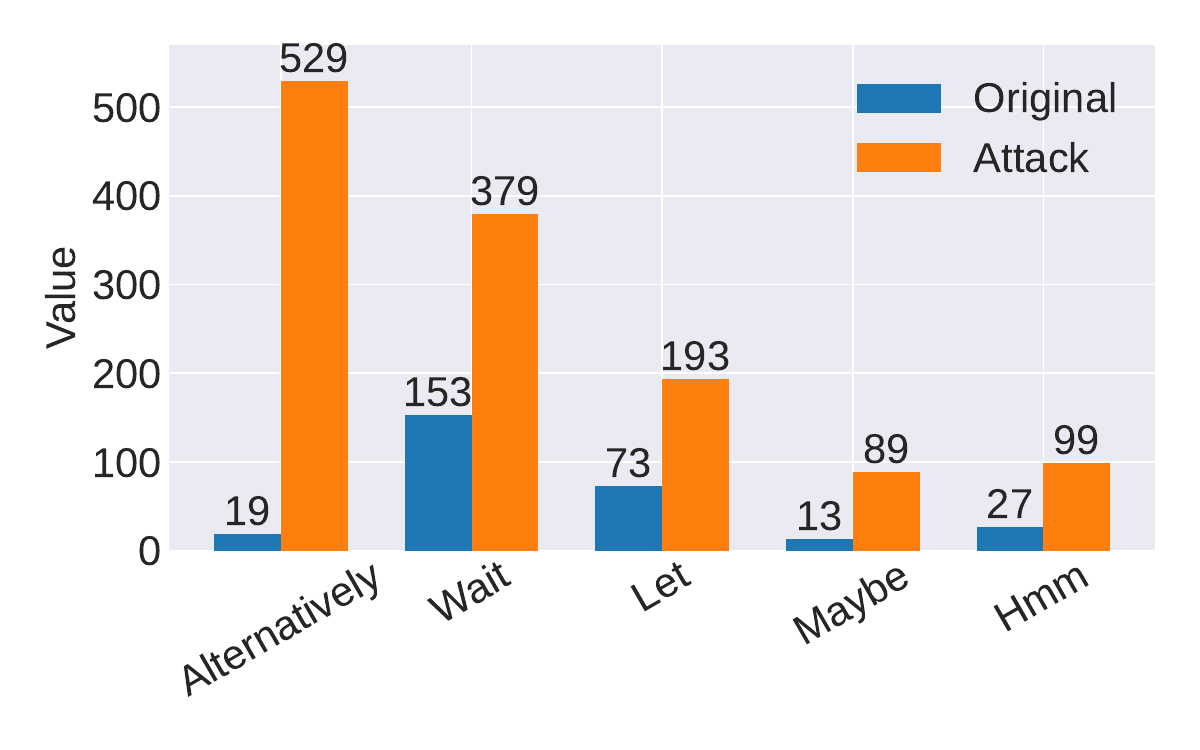}
    \caption{LLaMA}
\end{subfigure}
\begin{subfigure}{\columnwidth}
    \includegraphics[width=.9\linewidth]{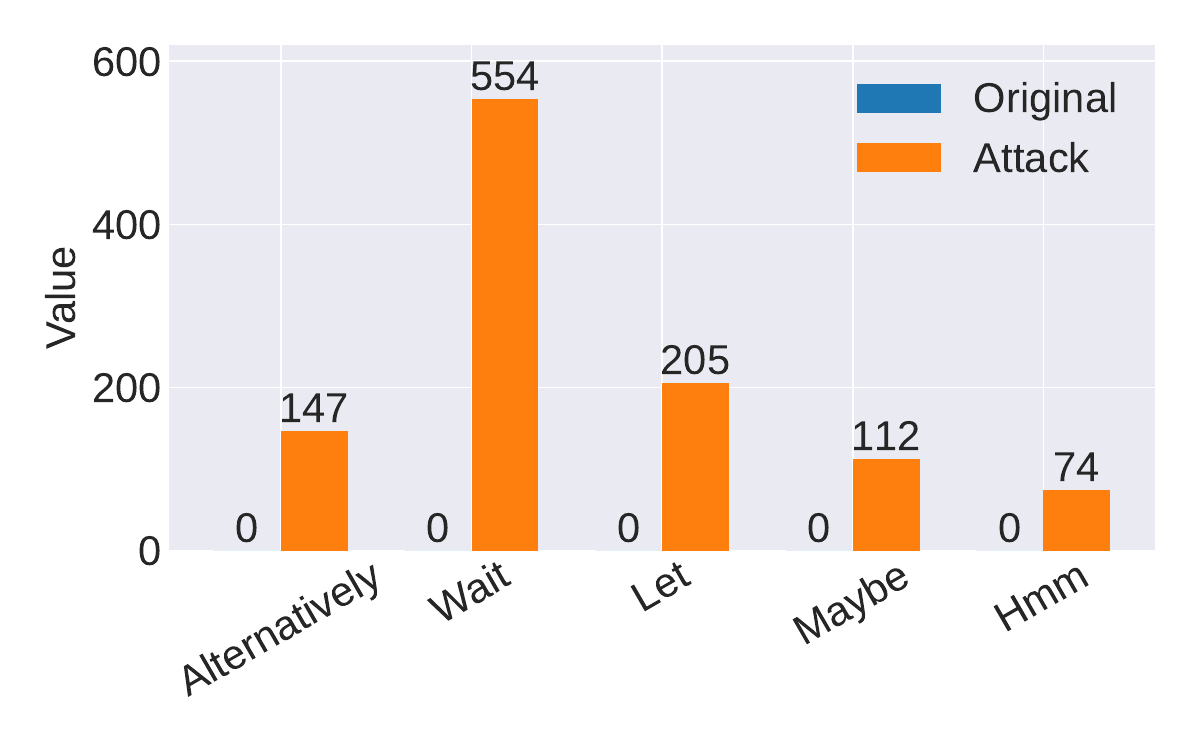}
    \caption{Qwen}
\end{subfigure}
\caption{Token counts between the generated outputs from the original and adversarial prompts.}
\label{fig:thinking_result}
\end{figure}

\mypara{Analysis}
To determine whether our adversarial suffixes truly elicit excessive reasoning rather than merely increasing output length, we conduct an analysis of the generated outputs.
First, we observe a substantial increase in the average number of reasoning sentences. 
For LLaMA, the average rises from 31 to 88, and for Qwen, from 7 to 74, when comparing outputs generated from original prompts to those generated with adversarial suffixes.
This pronounced increase suggests that our attack significantly extends the number of reasoning paths, rather than merely inflating output length.

Second, we analyze the distribution of the first two tokens in each reasoning sentence, comparing outputs generated with and without adversarial suffixes as shown in Fig.\ref{fig:thinking_result}.
The results reveal distinct lexical patterns between the two models.
For example, LLaMA more frequently uses deliberative tokens such as \textit{``Alternatively''} and \textit{``Wait''}, which are often associated with recursive reasoning.
In contrast, Qwen shows lower sensitivity to \textit{``Alternatively''}, suggesting that the expression of excessive reasoning may manifest differently across architectures.
Moreover, Qwen does not exhibit the same degree of excessive reasoning as LLaMA under standard conditions. 
However, it remains vulnerable to such behavior when exposed to adversarial suffixes.
Furthermore, the presence of tokens such as \textit{``Let''}, \textit{``Maybe''}, and \textit{``Hmm''}, which are difficult to detect through manual inspection, highlights the effectiveness of our ER Loss when combined with automated token selection.
This approach effectively surfaces subtle prompts capable of inducing excessive reasoning behavior.

\begin{table*}[t]
\centering
\resizebox{.9\linewidth}{!}{
\begin{tabular}{lcccc|cccc}
\toprule
 & \multicolumn{4}{c|}{LLaMA} & \multicolumn{4}{c}{Qwen} \\
 & Reason & Answer & Full & Accuracy & Reason & Answer & Full & Accuracy \\
 \midrule
o1-mini & 428 (+224) & 325 (+73) & 753 (+297) & 89\% (-2\%) & 591 (+386) & 512 (+261) & 1103 (+647) & 88\% (-3\%) \\
o3-mini & 446 (+199) & 184 (+46) & 630 (+245) & 90\% (+1\%) & 645 (+398) & 336 (+199) & 982 (+596) & 89\% (0\%) \\
R1 & 997 (-74) & 185 (-13) & 1182 (-87) & 95\% (-2\%) & 1295 (+224) & 233 (+36) & 1528 (+260) & 93\% (-4\%) \\
QWQ & 1761 (-151) & 231 (-6) & 1992 (-157) & 93\% (-3\%) & 2489 (+577) & 317 (+80) & 2806 (+657) & 98\% (+2\%) \\
\bottomrule
\end{tabular}
}
\caption{Transferability analysis of adversarial suffixes originally optimized for LLaMA and Qwen.
}
\label{tab:transfer_study}
\end{table*}

\mypara{Transferability}  
We evaluate the transferability of our adversarial suffixes to larger commercial language models, including o1-mini, o3-mini, DeepSeek-R1, and QWQ. 
Specifically, we test adversarial suffixes optimized on the LLaMA and Qwen models for the GSM8K dataset, with results summarized in Table~\ref{tab:transfer_study}.
Our findings show that these adversarial suffixes generalize effectively, consistently promoting longer output sequences without degrading task accuracy.
For the OpenAI model family, both LLaMA- and Qwen-optimized suffixes successfully increase output length. 
For example, suffixes optimized on LLaMA lead to a 245-token increase in total output length for o3-mini, and Qwen-optimized suffixes yield a 596-token increase.

In contrast, transferability to DeepSeek-R1 appears to depend on the source model.
Qwen-optimized suffixes result in a 260-token increase, whereas LLaMA-optimized suffixes fail to induce longer outputs. 
We hypothesize that this discrepancy is due to tokenizer compatibility, as DeepSeek-R1 shares the same tokenizer with Qwen but not with LLaMA.
A similar pattern is observed for QWQ, which also uses the Qwen tokenizer and shows greater sensitivity to Qwen-optimized suffixes.
These results suggest that while architectural differences influence the degree of computational overhead, tokenizer alignment plays a critical role in the transferability of adversarial prompts. 
Notably, Qwen appears to be a more effective proxy than LLaMA for crafting transferable adversarial suffixes across commercial systems.

\begin{figure}[t]
\centering
\begin{subfigure}{\columnwidth}
    \includegraphics[width=\linewidth]{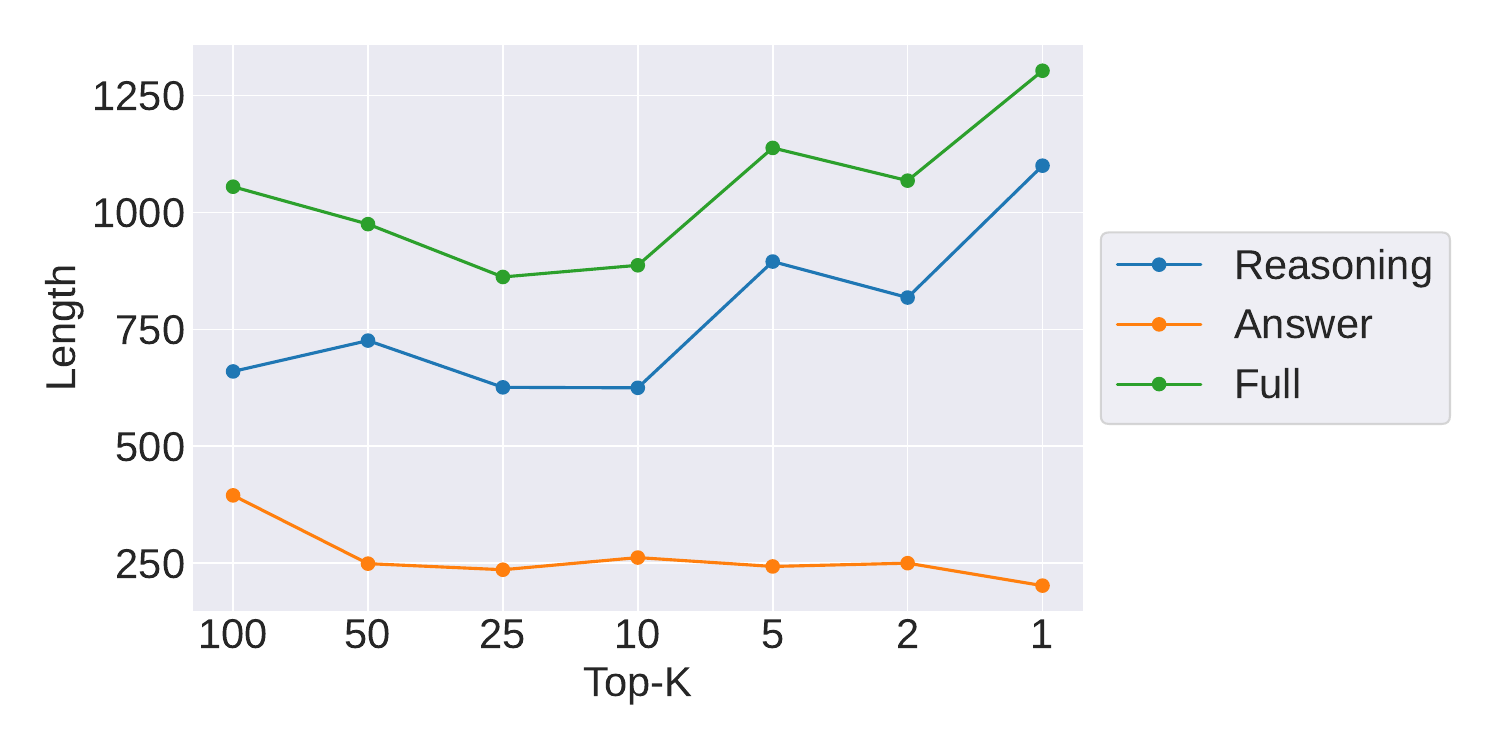}
\end{subfigure}
\caption{Impact of varying the top-K most informative tokens on LLaMA under greedy decoding.}
\label{fig:pce_study}
\end{figure}

\subsection{Ablation Studies}
\label{sec:ablation}

We conduct a series of ablation studies to assess the impact of different experimental configurations, including the introduction of the PCE Loss, the individual contribution of each loss component, and the effect of alternative target construction strategies.

\mypara{PCE Loss}
We begin by evaluating the effectiveness of the proposed PCE Loss by varying the proportion of top-$K$ tokens from 100\% to 1\%, as shown in Fig.~\ref{fig:pce_study}. 
The results show that focusing optimization on the top 5\%, and particularly the top 1\% of tokens, consistently outperforms applying loss uniformly across all tokens.
Peak performance is observed when focusing solely on the top 1\%, with the number of reasoning tokens increasing from 660 to 1,100.
This pattern suggests that selectively emphasizing a small subset of salient, prompt-dependent tokens can more effectively induce extended reasoning behavior. 
Additionally, we observe an inverse relationship between the number of reasoning and answer tokens, implying a redistribution of the model’s generative capacity toward reasoning content.
These findings underscore the value of targeted token optimization and demonstrate that prioritizing high-impact tokens is more effective than uniformly distributing the loss across the entire sequence.

\begin{table}[t]
\centering
\resizebox{\linewidth}{!}{
\begin{tabular}{lcccccc}
\toprule
Setup & Reason & Answer & Full & Latency & Energy & Accuracy \\
\midrule
$\mathcal{L}_{PCE}$ & 1100 & \textbf{202} & 1303 & 35.0 & 7016 & 84\% \\
$\mathcal{L}_{PCE} + \mathcal{L}_{ER}$ & 1169 & 188 & 1357 & 36.1 & 8089 & 88\% \\
$\mathcal{L}_{PCE} + \mathcal{L}_{DT}$ & 1447 & 201 & 1648 & 43.9 & 9558 & 88\% \\
$\mathcal{L}_{PCE} + \mathcal{L}_{ER} + \mathcal{L}_{DT}$ & \textbf{1914} & 160 & \textbf{2074} & \textbf{54.9} & \textbf{12827} & \textbf{92\%} \\
\bottomrule
\end{tabular}
}
\caption{Ablation study of loss objectives combinations on LLaMA under greedy decoding.}
\label{tab:loss_study}
\end{table}

\mypara{Loss Objectives}
Second, we evaluate the individual contributions of each loss function and assess their collective impact as presented in Table~\ref{tab:loss_study}. 
The results show that optimizing each loss independently leads to an increase in output sequence length, and the combination of all three loss functions yields the most substantial gains in both sequence length and computational burden. 
For instance, the full composite loss achieves the longest average output (up to 2,074 tokens), the highest inference latency (54.9 seconds), and the greatest energy consumption (12,827J). 
These results underscore the synergistic effect of combining all three objectives.

To further analyze the behavior encouraged by the ER Loss, we visualize a word cloud of the most frequently prioritized tokens in Fig.~\ref{fig:wordcloud_result}. 
Common deliberative tokens identified in prior work, such as \textit{``Alternatively''} and \textit{``Wait''}, are prominently featured. 
In addition, our method surfaces previously underexplored tokens such as \textit{``Maybe''} and \textit{``Hmm''}, which act as effective triggers for extended reasoning.
These findings confirm that the joint loss formulation effectively amplifies reasoning behavior while preserving task accuracy, and that the ER Loss successfully uncovers subtle lexical cues indicative of recursive reasoning.

\begin{table}[t]
\centering
\resizebox{\linewidth}{!}{
\begin{tabular}{lcccccc}
\toprule
Setup & Reason & Answer & Full & Latency & Energy & Accuracy \\
\midrule
Raw & 1695 & 190 & 1885 & 50.9 & 12120 & 80\% \\
CoT & 1060 & \textbf{224} & 1283 & 34.0 & 7719 & 80\% \\
CoT + DSPy & \textbf{1914} & 160 & \textbf{2074} & \textbf{54.9} & \textbf{12827} & \textbf{92\%} \\
\bottomrule
\end{tabular}
}
\caption{Ablation study of different target constructions with our proposed loss function on LLaMA under greedy decoding.}
\label{tab:target_study}
\end{table}

\mypara{Target Output Construction}
Finally, we evaluate several strategies for constructing target outputs to guide adversarial optimization, as summarized in Table~\ref{tab:target_study}. 
The comparison includes a raw baseline (no additional prompt), a standard CoT prompt, and a DSPy-optimized CoT prompt.
Interestingly, we find that the standard CoT prompt does not consistently produce longer reasoning sequences; in some cases, it even results in shorter outputs than raw prompting, highlighting its limitations in eliciting extended reasoning.
In contrast, DSPy-optimized CoT prompts increase the average output length from 1,283 to 2,074 tokens under greedy decoding compared to CoT prompts, with corresponding increases in both energy consumption and task accuracy.
These results highlight the critical role of target output quality in guiding adversarial optimization. 
Longer reasoning sequences, especially those produced via DSPy, serve as more effective targets for inducing excessive computation. 
This reinforces the importance of target construction in maximizing the efficacy of our attack.

\section{Conclusion and Discussion}
\label{sec:conclusion}

In this work, we present a novel adversarial attack targeting reasoning LLMs by inducing significant computational overhead during inference. 
Our approach constructs adversarial suffixes that trigger extended reasoning trajectories, guided by a composite loss function tailored to maximize output length and complexity.
Empirical evaluations show that our method consistently increases output sequence length, inference latency, and energy consumption, without compromising task performance. 
Furthermore, our results reveal strong cross-model transferability, underscoring the practical relevance of this threat in real-world settings.

\mypara{Limitations and Future Work}
Despite its effectiveness, our current framework assumes white-box access to model internals, including gradients and token embeddings. 
Nevertheless, we demonstrate partial transferability to several commercial, black-box LLMs, suggesting potential for broader applicability.
Future work will explore black-box optimization strategies to extend the attack to restricted-access models. 
Additionally, our current experiments are limited to target sequences of up to 3,000 tokens due to computational constraints. 
Scaling to longer contexts is a promising direction for further study.

\section*{Broader Impact}
\label{sec:impact}

This study uncovers a new class of inference-time vulnerabilities in reasoning LLMs, with significant implications for both computational security and resource efficiency. 
We demonstrate that minimal adversarial perturbations, crafted as short suffixes, can dramatically increase inference costs. 
In large-scale deployments, such inefficiencies may translate into substantial computational burdens.

Moreover, the transferability of these adversarial behaviors across model families highlights a pressing concern for the robustness and reliability of LLMs in real-world applications. 
These results call for the development of evaluation frameworks that go beyond accuracy to include efficiency, resilience, and scalability under adversarial conditions.

We advocate for the integration of resource-awareness and adversarial robustness into the design and deployment of future LLMs to ensure their sustainability and operational safety. 
To support continued research in this direction, we will release our codebase publicly, contributing to the development of more secure and efficient language model systems.

\begin{small}
\bibliographystyle{plain}
\bibliography{normal_generated_py3}

\begin{thebibliography}{10}

\bibitem{huggingface}
\url{https://huggingface.co/}.

\bibitem{BBKGPGGLNNH24}
Maciej Besta, Nils Blach, Ales Kubicek, Robert Gerstenberger, Michal Podstawski, Lukas Gianinazzi, Joanna Gajda, Tomasz Lehmann, Hubert Niewiadomski, Piotr Nyczyk, and Torsten Hoefler.
\newblock {Graph of Thoughts: Solving Elaborate Problems with Large Language Models}.
\newblock In {\em {AAAI Conference on Artificial Intelligence (AAAI)}}, pages 17682--17690. AAAI, 2024.

\bibitem{BSAP22}
Nicholas Boucher, Ilia Shumailov, Ross Anderson, and Nicolas Papernot.
\newblock {Bad Characters: Imperceptible {NLP} Attacks}.
\newblock In {\em {IEEE Symposium on Security and Privacy (S\&P)}}, pages 1987--2004. IEEE, 2022.

\bibitem{CNCJGKITS23}
Nicholas Carlini, Milad Nasr, Christopher~A. Choquette{-}Choo, Matthew Jagielski, Irena Gao, Pang~Wei Koh, Daphne Ippolito, Florian Tram{\`{e}}r, and Ludwig Schmidt.
\newblock {Are aligned neural networks adversarially aligned?}
\newblock In {\em {Annual Conference on Neural Information Processing Systems (NeurIPS)}}. NeurIPS, 2023.

\bibitem{CSHLY22}
Simin Chen, Zihe Song, Mirazul Haque, Cong Liu, and Wei Yang.
\newblock {NICGSlowDown: Evaluating the Efficiency Robustness of Neural Image Caption Generation Models}.
\newblock In {\em {IEEE Conference on Computer Vision and Pattern Recognition (CVPR)}}, pages 15344--15353. IEEE, 2022.

\bibitem{CXLHPYSLZZWTMY24}
Xingyu Chen, Jiahao Xu, Tian Liang, Zhiwei He, Jianhui Pang, Dian Yu, Linfeng Song, Qiuzhi Liu, Mengfei Zhou, Zhuosheng Zhang, Rui Wang, Zhaopeng Tu, Haitao Mi, and Dong Yu.
\newblock {Do {NOT} Think That Much for 2+3=? On the Overthinking of o1-Like LLMs}.
\newblock {\em {CoRR abs/2412.21187}}, 2024.

\bibitem{CKBCJKPTHNHS21}
Karl Cobbe, Vineet Kosaraju, Mohammad Bavarian, Mark Chen, Heewoo Jun, Lukasz Kaiser, Matthias Plappert, Jerry Tworek, Jacob Hilton, Reiichiro Nakano, Christopher Hesse, and John Schulman.
\newblock {Training Verifiers to Solve Math Word Problems}.
\newblock {\em {CoRR abs/2110.14168}}, 2021.

\bibitem{DGYZSZXZMWBZYWWGSLGLXWWFLZDZRDCJLLDLHCLZBXWDXGQLGLWCYQLCNLCDHGGHYWZZWZXXZZTLWLTHZWCDGZPWCJCLZCYWYZPL25}
DeepSeek{-}AI, Daya Guo, Dejian Yang, Haowei Zhang, Junxiao Song, Ruoyu Zhang, Runxin Xu, Qihao Zhu, Shirong Ma, Peiyi Wang, Xiao Bi, Xiaokang Zhang, Xingkai Yu, Yu~Wu, Z.~F. Wu, Zhibin Gou, Zhihong Shao, Zhuoshu Li, Ziyi Gao, Aixin Liu, Bing Xue, Bingxuan Wang, Bochao Wu, Bei Feng, Chengda Lu, Chenggang Zhao, Chengqi Deng, Chenyu Zhang, Chong Ruan, Damai Dai, Deli Chen, Dongjie Ji, Erhang Li, Fangyun Lin, Fucong Dai, Fuli Luo, Guangbo Hao, Guanting Chen, Guowei Li, H.~Zhang, Han Bao, Hanwei Xu, Haocheng Wang, Honghui Ding, Huajian Xin, Huazuo Gao, Hui Qu, Hui Li, Jianzhong Guo, Jiashi Li, Jiawei Wang, Jingchang Chen, Jingyang Yuan, Junjie Qiu, Junlong Li, J.~L. Cai, Jiaqi Ni, Jian Liang, Jin Chen, Kai Dong, Kai Hu, Kaige Gao, Kang Guan, Kexin Huang, Kuai Yu, Lean Wang, Lecong Zhang, Liang Zhao, Litong Wang, Liyue Zhang, Lei Xu, Leyi Xia, Mingchuan Zhang, Minghua Zhang, Minghui Tang, Meng Li, Miaojun Wang, Mingming Li, Ning Tian, Panpan Huang, Peng Zhang, Qiancheng Wang, Qinyu Chen, Qiushi Du, Ruiqi Ge,
  Ruisong Zhang, Ruizhe Pan, Runji Wang, R.~J. Chen, R.~L. Jin, Ruyi Chen, Shanghao Lu, Shangyan Zhou, Shanhuang Chen, Shengfeng Ye, Shiyu Wang, Shuiping Yu, Shunfeng Zhou, Shuting Pan, and S.~S. Li.
\newblock {DeepSeek-R1: Incentivizing Reasoning Capability in LLMs via Reinforcement Learning}.
\newblock {\em {CoRR abs/2501.12948}}, 2025.

\bibitem{GZCMGKKS24}
Esther Gan, Yiran Zhao, Liying Cheng, Yancan Mao, Anirudh Goyal, Kenji Kawaguchi, Min{-}Yen Kan, and Michael Shieh.
\newblock {Reasoning Robustness of LLMs to Adversarial Typographical Errors}.
\newblock In {\em {Conference on Empirical Methods in Natural Language Processing (EMNLP)}}, pages 10449--10459. ACL, 2024.

\bibitem{GBGXTLL24}
Kuofeng Gao, Yang Bai, Jindong Gu, Shu{-}Tao Xia, Philip Torr, Zhifeng Li, and Wei Liu.
\newblock {Inducing High Energy-Latency of Large Vision-Language Models with Verbose Images}.
\newblock In {\em {International Conference on Learning Representations (ICLR)}}, 2024.

\bibitem{JKLRELHMBCIKPNPWTBKSVDKMAJNZGRSBMMHBHMELBHLBVSZKOHFCRKLSDFMRTLOFZWPCMWRMWSRLTSLSSPCZBLSBSiBRLCKOOGA24}
Aaron Jaech, Adam Kalai, Adam Lerer, Adam Richardson, Ahmed El{-}Kishky, Aiden Low, Alec Helyar, Aleksander Madry, Alex Beutel, Alex Carney, Alex Iftimie, Alex Karpenko, Alex~Tachard Passos, Alexander Neitz, Alexander Prokofiev, Alexander Wei, Allison Tam, Ally Bennett, Ananya Kumar, Andre Saraiva, Andrea Vallone, Andrew Duberstein, Andrew Kondrich, Andrey Mishchenko, Andy Applebaum, Angela Jiang, Ashvin Nair, Barret Zoph, Behrooz Ghorbani, Ben Rossen, Benjamin Sokolowsky, Boaz Barak, Bob McGrew, Borys Minaiev, Botao Hao, Bowen Baker, Brandon Houghton, Brandon McKinzie, Brydon Eastman, Camillo Lugaresi, Cary Bassin, Cary Hudson, Chak~Ming Li, Charles de~Bourcy, Chelsea Voss, Chen Shen, Chong Zhang, Chris Koch, Chris Orsinger, Christopher Hesse, Claudia Fischer, Clive Chan, Dan Roberts, Daniel Kappler, Daniel Levy, Daniel Selsam, David Dohan, David Farhi, David Mely, David Robinson, Dimitris Tsipras, Doug Li, Dragos Oprica, Eben Freeman, Eddie Zhang, Edmund Wong, Elizabeth Proehl, Enoch Cheung, Eric Mitchell,
  Eric Wallace, Erik Ritter, Evan Mays, Fan Wang, Felipe~Petroski Such, Filippo Raso, Florencia Leoni, Foivos Tsimpourlas, Francis Song, Fred von Lohmann, Freddie Sulit, Geoff Salmon, Giambattista Parascandolo, Gildas Chabot, Grace Zhao, Greg Brockman, Guillaume Leclerc, Hadi Salman, Haiming Bao, Hao Sheng, Hart, in, Hessam Bagherinezhad, Hongyu Ren, Hunter Lightman, Hyung~Won Chung, Ian Kivlichan, Ian O'Connell, Ian Osband, Ignasi~Clavera Gilaberte, and Ilge Akkaya.
\newblock {OpenAI o1 System Card}.
\newblock {\em {CoRR abs/2412.16720}}, 2024.

\bibitem{KSMZSVHSJMMZP24}
Omar Khattab, Arnav Singhvi, Paridhi Maheshwari, Zhiyuan Zhang, Keshav Santhanam, Sri Vardhamanan, Saiful Haq, Ashutosh Sharma, Thomas~T. Joshi, Hanna Moazam, Heather Miller, Matei Zaharia, and Christopher Potts.
\newblock {DSPy: Compiling Declarative Language Model Calls into State-of-the-Art Pipelines}.
\newblock In {\em {International Conference on Learning Representations (ICLR)}}, 2024.

\bibitem{KRNKIHB25}
Abhinav Kumar, Jaechul Roh, Ali Naseh, Marzena Karpinska, Mohit Iyyer, Amir Houmansadr, and Eugene Bagdasarian.
\newblock {OverThink: Slowdown Attacks on Reasoning LLMs}.
\newblock {\em {CoRR abs/2502.02542}}, 2025.

\bibitem{MKRA24}
Arindam Mitra, Hamed Khanpour, Corby Rosset, and Ahmed Awadallah.
\newblock {Orca-Math: Unlocking the potential of SLMs in Grade School Math}.
\newblock {\em {CoRR abs/2402.14830}}, 2024.

\bibitem{RRTYBMZR25}
Meghana~Arakkal Rajeev, Rajkumar Ramamurthy, Prapti Trivedi, Vikas Yadav, Oluwanifemi Bamgbose, Sathwik~Tejaswi Madhusudhan, James Zou, and Nazneen Rajani.
\newblock {Cats Confuse Reasoning {LLM:} Query Agnostic Adversarial Triggers for Reasoning Models}.
\newblock {\em {CoRR abs/2503.01781}}, 2025.

\bibitem{SZBPMA21}
Ilia Shumailov, Yiren Zhao, Daniel Bates, Nicolas Papernot, Robert~D. Mullins, and Ross Anderson.
\newblock {Sponge Examples: Energy-Latency Attacks on Neural Networks}.
\newblock In {\em {IEEE European Symposium on Security and Privacy (Euro S\&P)}}, pages 212--231. IEEE, 2021.

\bibitem{qwq32b}
Qwen Team.
\newblock {QwQ-32B: Embracing the Power of Reinforcement Learning}.
\newblock \url{https://qwenlm.github.io/blog/qwq-32b/}, 2025.

\bibitem{WWSLCNCZ23}
Xuezhi Wang, Jason Wei, Dale Schuurmans, Quoc~V. Le, Ed~H. Chi, Sharan Narang, Aakanksha Chowdhery, and Denny Zhou.
\newblock {Self-Consistency Improves Chain of Thought Reasoning in Language Models}.
\newblock In {\em {International Conference on Learning Representations (ICLR)}}, 2023.

\bibitem{WLXLCHSYLZWTMY25}
Yue Wang, Qiuzhi Liu, Jiahao Xu, Tian Liang, Xingyu Chen, Zhiwei He, Linfeng Song, Dian Yu, Juntao Li, Zhuosheng Zhang, Rui Wang, Zhaopeng Tu, Haitao Mi, and Dong Yu.
\newblock {Thoughts Are All Over the Place: On the Underthinking of o1-Like LLMs}.
\newblock {\em {CoRR abs/2501.18585}}, 2025.

\bibitem{WWSBIXCLZ22}
Jason Wei, Xuezhi Wang, Dale Schuurmans, Maarten Bosma, Brian Ichter, Fei Xia, Ed~H. Chi, Quoc~V. Le, and Denny Zhou.
\newblock {Chain-of-Thought Prompting Elicits Reasoning in Large Language Models}.
\newblock In {\em {Annual Conference on Neural Information Processing Systems (NeurIPS)}}. NeurIPS, 2022.

\bibitem{YYZCXZGC24}
Ling Yang, Zhaochen Yu, Tianjun Zhang, Shiyi Cao, Minkai Xu, Wentao Zhang, Joseph~E. Gonzalez, and Bin Cui.
\newblock {Buffer of Thoughts: Thought-Augmented Reasoning with Large Language Models}.
\newblock {\em {CoRR abs/2406.04271}}, 2024.

\bibitem{ZWKF23}
Andy Zou, Zifan Wang, J.~Zico Kolter, and Matt Fredrikson.
\newblock {Universal and Transferable Adversarial Attacks on Aligned Language Models}.
\newblock {\em {CoRR abs/2307.15043}}, 2023.

\end{thebibliography}
\end{small}

\clearpage
\appendix

\section{Additional Experimental Details}

In this section, we present additional details of our experimental setup, including the prompting templates, analysis of the attack behavior, descriptions of the models and datasets employed, as well as the computational resources utilized.

\subsection{Prompt Template}

\tcbset{colback=gray!5!white, colframe=gray!80!black, 
    width=\textwidth, boxrule=0.5mm, arc=2mm, auto outer arc, 
    left=1mm, right=1mm, top=1mm, bottom=1mm}
\begin{center}
\begin{cvbox}[~~Optimized CoT prompt using DSPy]

Provide a detailed breakdown of the problem or topic at hand, considering every aspect and nuance. Start from an introductory overview, followed by dissecting the components logically, integrating examples, addressing possible counterarguments or misconceptions, and uniformly elaborating on each point with explanatory details. Prioritize presenting a thorough, analytical approach that can include speculative insights, hypothetical scenarios, or alternative viewpoints, without rushing to conclusions. Aim to showcase a clear conceptualization followed by logical reasoning, actively prompting self-reflection on the implications of your insights.

\end{cvbox}
\label{box:dspy_prompt}
\end{center}

\tcbset{colback=gray!5!white, colframe=gray!80!black, 
    width=\textwidth, boxrule=0.5mm, arc=2mm, auto outer arc, 
    left=1mm, right=1mm, top=1mm, bottom=1mm}
\begin{center}
\begin{cvbox}[~~Prompting template for extracting answer]

Here is a math question and a model's answer about this question. 

Please extract the EXACT number from the answer text as the final answer for question.

QUESTION: \{\}

ANSWER: \{\}

Final format should be a legal 'number' without any suffix such as '\$'.

The final answer is:

\end{cvbox}
\label{box:acc_prompt}
\end{center}

\clearpage

\subsection{Experimental Setups}
\label{sec:resouces}

All models in our experiments are downloaded from HuggingFace. 
DeepSeek-R1-Distill-Llama-8B and DeepSeek-R1-Distill-Qwen-7B are originally licensed under the Apache 2.0 License.
The GSM8K and ORCA datasets are under the MIT License.
All experiments in the paper were conducted on an A100 (80GB) compute node.

\subsection{Results \& Statistics}
\label{sec:add_examples}

\begin{figure}[h]
\centering
\begin{subfigure}{\columnwidth}
    \includegraphics[width=\linewidth]{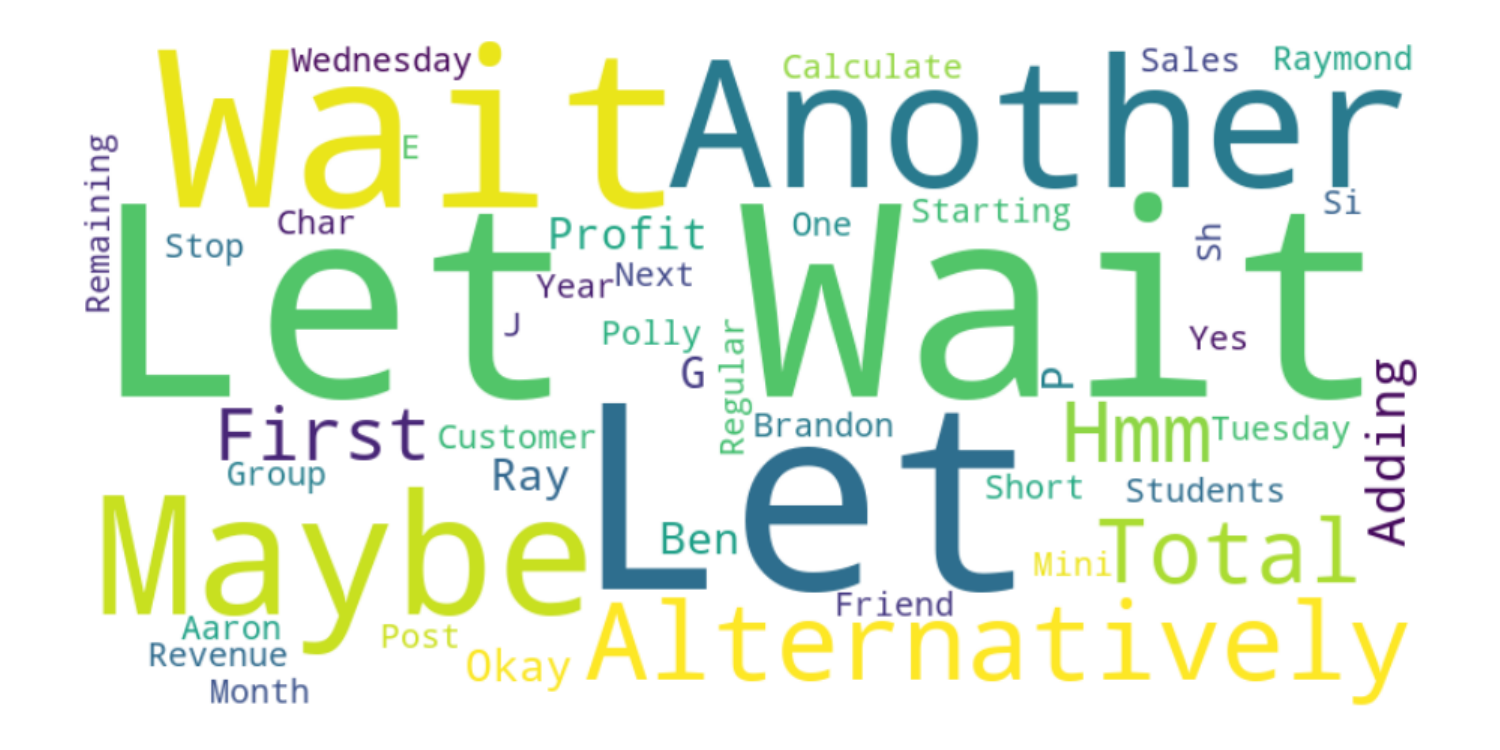}
    \caption{LLaMA}
\end{subfigure}
\begin{subfigure}{\columnwidth}
    \includegraphics[width=\linewidth]{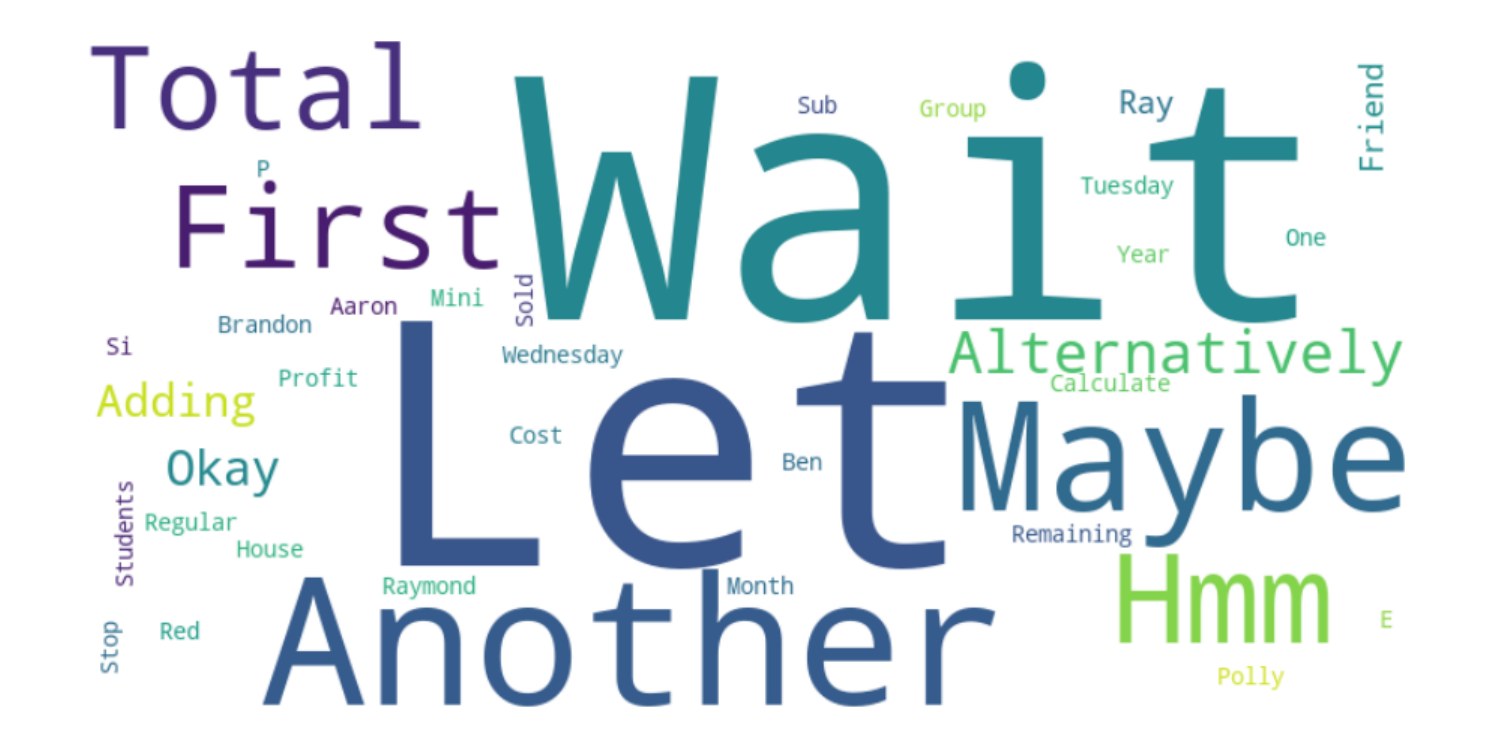}
    \caption{Qwen}
\end{subfigure}
\caption{Tokens used for Trajectory Expansion Loss. 
Word clouds generated from the CoT outputs of LLaMA and Qwen on GSM8K.}
\label{fig:wordcloud_result}
\end{figure}

\begin{figure}[t]
\centering
\begin{subfigure}{\columnwidth}
    \includegraphics[width=\linewidth]{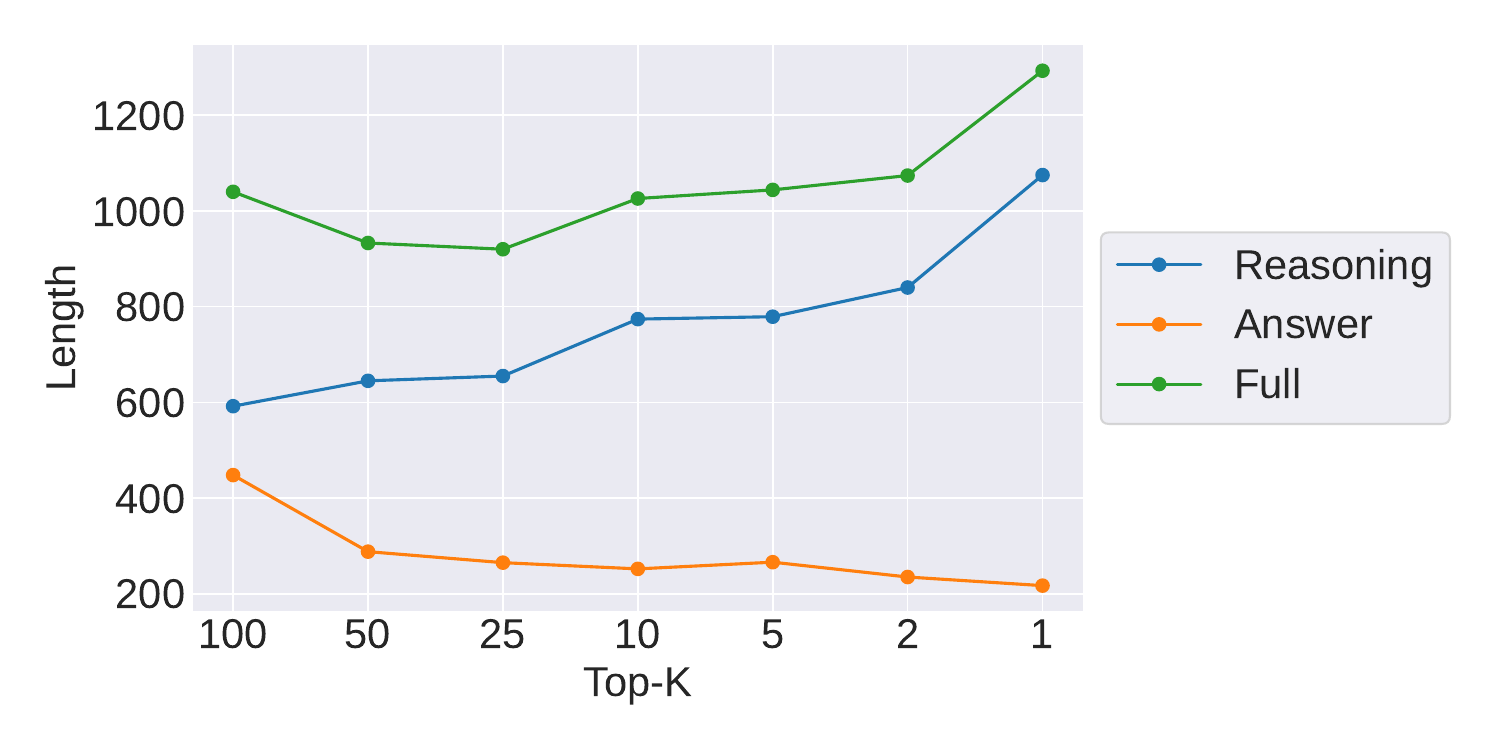}
\end{subfigure}
\caption{Impact of varying the top-K most informative tokens on LLaMA under sampling decoding.}
\label{fig:pce_study_sample}
\end{figure}

\begin{table}[h]
\centering
\begin{tabular}{lc}
\toprule
 & Mean \\
 \midrule
Raw & 805 \\
CoT & 717 \\
CoT+DSPy & 1247 \\
\bottomrule
\end{tabular}
\caption{The statistics of sample outputs with different promptings.}
\label{tab:target_stat}
\end{table}

\begin{table}[h]
\centering
\resizebox{\linewidth}{!}{
\begin{tabular}{lcccccc}
\toprule
Setup & Reason & Answer & Full & Latency & Energy & Accuracy \\
\midrule
$\mathcal{L}_{PCE}$ & 1104 & 218 & 1322 & 143.9 & 19810 & 86\% \\
$\mathcal{L}_{PCE} + \mathcal{L}_{ER}$ & 1095 & 205 & 1301 & 124.9 & 14879 & 87\% \\
$\mathcal{L}_{PCE} + \mathcal{L}_{DT}$ & 1184 & \textbf{223} & 1407 & 145.3 & 17731 & 88\% \\
$\mathcal{L}_{PCE} + \mathcal{L}_{ER} + \mathcal{L}_{DT}$ & \textbf{1437} & 204 & \textbf{1641} & \textbf{197.0} & \textbf{21228} & \textbf{90\%} \\
\bottomrule
\end{tabular}
}
\caption{Ablation study of loss objectives combinations on LLaMA under sampling decoding.}
\label{tab:loss_study_sample}
\end{table}

\begin{table}[h]
\centering
\resizebox{\linewidth}{!}{
\begin{tabular}{lcccccc}
\toprule
Setup & Reason & Answer & Full & Latency & Energy & Accuracy \\
\midrule
Raw & 1396 & 223 & 1619 & 206.8 & \textbf{20252} & 82\% \\
CoT & 1034 & \textbf{250} & 1283 & 150.2 & 15685 & 79\% \\
CoT + DSPy & \textbf{1501} & 216 & \textbf{1718} & \textbf{209.9} & 19491 & \textbf{87\%} \\
\bottomrule
\end{tabular}
}
\caption{Ablation study of different target construction strategies on LLaMA under sampling decoding.}
\label{tab:target_study_sample}
\end{table}

\end{document}